\documentclass[onecolumn]{aa}
\usepackage{graphicx}
\usepackage{txfonts}
\newcommand{\etal}{{\it et al. }}
\newcommand{\src}{XMMU~J004308.6+411247}
\newcommand{\R}{{\em ROSAT}}
\newcommand{\X}{XMM--{\em Newton}}
\newcommand{\C}{{\em Chandra}}
\newcommand{\B}{{\em Beppo}--SAX}
\def\ltsima{$\; \buildrel < \over \sim \;$}
\def\simlt{\lower.5ex\hbox{\ltsima}}
\def\gtsima{$\; \buildrel > \over \sim \;$}
\def\simgt{\lower.5ex\hbox{\gtsima}}
\begin{document}
\title{A possible new dipping X--ray source in the field of M31}
\author{Vanessa Mangano\inst{1}, 
Gian Luca Israel\inst{1} 
and Luigi Stella\inst{1}} 
\institute{
$^{1}$ Osservatorio Astronomico di Roma, 
Via Frascati 33, I--00040 Monteporzio Catone (Roma), Italy 
}
\date{Submitted ..... / Accepted .....}
\offprints{V. Mangano, mangano@mporzio.astro.it}
\abstract {
We report the discovery of the new dipping  X--ray source, \src, in M31,
during a systematic search for periodicities in \X\ archival observations.
During the 2002 January 6 observation, 
the dips recur with a 107\,min period and 
the source count rate is consistent with zero 
at the dip minimum. 
Dips with the same modulation and period are also observed 
during the \X\ observations carried out on 2000 June 25, 2001 June 29 
and the \C\ observation of 2001 October 5. 
The dips of \src\ show no evidence of energy dependence. 
The average X--ray flux of \src\ is 
nearly constant across different observations
($\sim 10^{37}$\,erg\,s$^{-1}$ for an assumed M31 distance of 780\,kpc 
in the 0.3--10\,keV band); the spectrum is 
well fit by an absorbed power law with a photon index $\sim 0.8$  
or an absorbed Comptonization model. 
The photo-electric absorption is consistent with 
the Galactic value in the source direction.
If \src\ is located in M31 its properties are
consistent with those of dipping low mass X--ray binaries in the Galaxy.
Present observations do not allow to distinguish between dips and eclipses.

The possibility that \src\ is a foreground X--ray source cannot be ruled out
at present; 
in this case the source might be a magnetic cataclysmic variable.  

}

\authorrunning{Mangano, V., Israel, G.L., Stella, L.}
\titlerunning{A possible new dipping X--ray source in M31}
\maketitle

\section{Introduction}
\label{intro}

Being the closest large spiral (Sb) galaxy to our own, 
the Andromeda Galaxy (M31) provides us with a prime
opportunity to study X--ray emission in a Galaxy similar 
to the Milky Way both in morphology and size.
The sources in M31 are observed 
at the nearly uniform and well known
distance of $\sim$780\,kpc 
(Stanek \& Garnavich \cite{staga98}; Macri \etal \cite{macri01}).
Moreover, owing to its inclination angle of $\sim 77^{\circ}$ 
and high galactic latitude ($l$II$=121.174312,\;b$II$=-21.573022$)
sources are viewed through a substantially lower absorption column 
($N_{\rm H}\sim 7\times 10^{20}$\,cm$^2$, Dickey \& Lockman \cite{dilo90})
than for sources in the Galaxy. 
Because of this moderate and fairly uniform extinction, M31 sources
can be studied and compared over an extended low energy band. 
Moreover, different stellar populations in the bulge, spiral arms
and halo can be easily distinguished based on their location. 
The \X\ mission (Jansen \etal \cite{jansen01})
yields an unprecedented opportunity to study the time variability
and spectral properties of individual X--ray sources in nearby galaxies
with increased throughput.
Previous observations of M31 
performed with {\it Einstein} (Trinchieri \& Fabbiano \cite{trifab91}; 
van Speybroeck \etal \cite{vanspey79}) 
and \R\ (Primini, Forman \& Jones \cite{prifojo93}) 
revealed many point-like X--ray sources in its core
($r < 5^{\prime} \sim 1.1$\,kpc),
but were too insensitive for their precise characterization.
The two deep and extensive \R--{\it PSPC} surveys
of the entire M31 disk (Supper \etal \cite{supper01}; 
Supper \etal \cite{supper97}), though 
effective in mapping the Globular Cluster (GC) 
and Supernova Remnant (SNR) populations, 
proved inadequate for a detailed study of
point sources in
the core of the galaxy
because of the limited spatial resolution.
Spatial resolution was the main limitation
also in the study of the bulge of M31 carried out with \B,
the first to image M31 up to energies of $\sim$10\,keV 
(Trinchieri \etal \cite{trinch99}).

The observations of the central part of M31 with the \C\ 
X--Ray Observatory (Weisskopf \cite{weiss88})
resolved the nuclear source seen
with the {\it Einstein} and \R--{\it HRI} 
into five point sources (Garcia \etal \cite{garcia00})
and detected and localized
bulge point sources down to luminosities
of $\sim 2 \times 10^{35}$\,erg\,s$^{-1}$  
in the 0.3--7\,keV band,
with positional errors of less than 1$^{\prime\prime}$
(Kong \etal \cite{kong02}; Di Stefano \etal \cite{diste02}; 
Kaaret \cite{kaaret02}).
None of these observations, however, was long enough 
for a detailed timing analysis.
The large collecting area and bandpass of \X\ 
afforded a more accurate characterization
of both the global properties of the X--ray emission from M31
(Shirey \etal \cite{shirey01}) 
and individual X--ray sources in its bulge 
(Osborne \etal \cite{osborne01}; Barnard, Kolb \& Osborne \cite{bakosb02}; 
Trudolyubov \etal \cite{trudo02}).
M31 was selected
as an \X\ Performance Verification target 
(Jansen \etal \cite{jansen01}) and subsequently observed 
within the Guaranteed Time Program.
The \X\ observations of M31 
(see section \ref{obs&data})
are among the M31 X--ray observations with the longest total exposure; 
they are eminently suited for timing and spectral analysis.
We started a program of systematic search
for periodicities with Fourier techniques among all bright 
(\simgt $5 \times 10^{35}$\,erg\,s$^{-1}$) point-like X--ray sources detected 
in the \X\ archival observations of M31.
Our search strategy is aimed at detecting periodic signals 
over the widest possible period range 
(see section \ref{strategy}).
In this paper
we report the discovery of a 107\,min dip-like modulation 
(nearly 100\% amplitude)
in the 2002 January 6 light curve of the source \src.
The same modulation is clearly seen 
also in other \X\ observations of M31  
and in the ACIS--S \C\ observation of 2001 October 5 
(see section \ref{res-timingan}).

In the following sections 
we describe the set of data we analysed (section \ref{obs&data}),
summarise the periodicity search technique (section \ref{strategy})
and illustrate the properties of the dip we discovered 
in the light curve of \src\ (section \ref{results}). 
Our results are discussed in section \ref{discussion}.
Throughout the paper we adopt a distance to M31 of 780\,kpc 
(Stanek \& Garnavich \cite{staga98}; Macri \etal \cite{macri01})
and a Galactic column density 
in the direction of M31 of $7\times 10^{20}$\,cm$^{-2}$
(Dickey \& Lockman \cite{dilo90}).


\section{Observations and data analysis}
\label{obs&data}

\begin{figure}
\centering
\includegraphics[width=12cm]{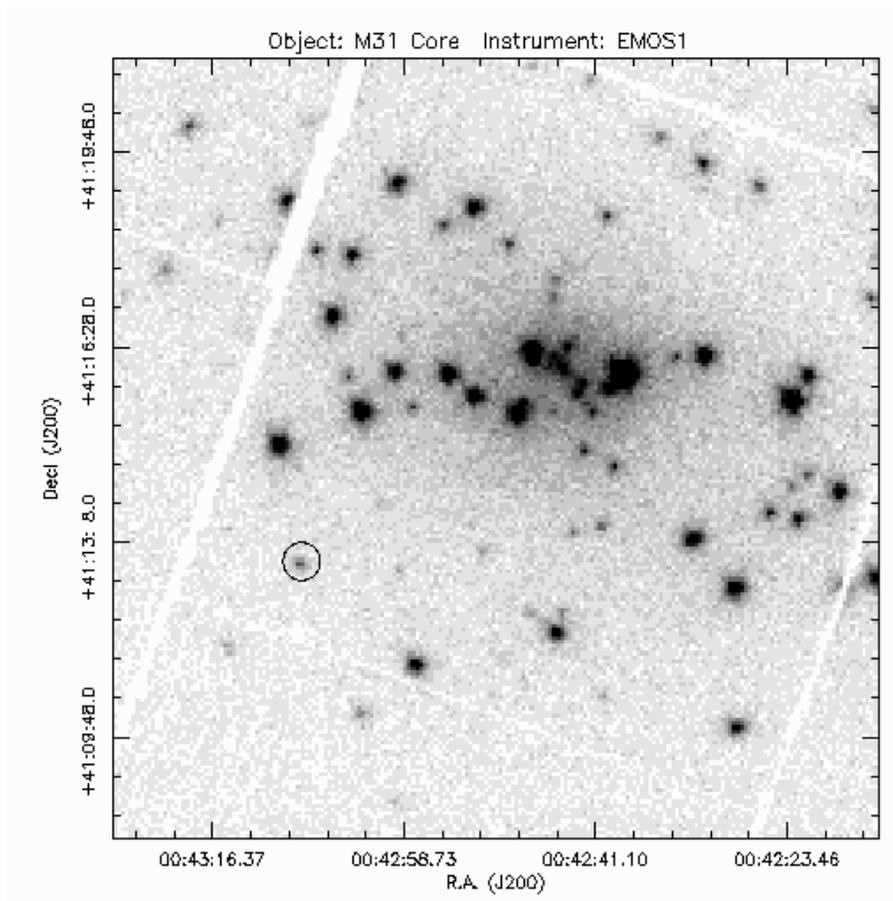}
\caption{Image of M31 from the \X\ observation carried out on 
2002 January 6 (MOS1 data). 
The source \src\ is marked by the extraction circle that we used, 
containing 98\% of the source counts.
 \label{m31image}}
\end{figure}

We used the four archival \X\ observations
of the core of M31 listed in Table \ref{datatab}.
The 2000 June 25 observation 
(observation 1 in Table \ref{datatab}) 
was part of the Performance 
Verification Program (Jansen \etal \cite{jansen01}; 
Shirey \etal \cite{shirey01}; Osborne \etal \cite{osborne01}).
The other three observations (observations 2,3,4) 
were carried out on
2000 December 28, 2001 June 29 and 2002 January 6 respectively 
as a part of Guaranteed Time Program.
We used data from the three European Photon Imaging Camera
(EPIC) instruments at the focus of the three co--aligned 1500\,cm$^2$ 
X--ray telescopes on board \X:
two EPIC MOS detectors (Turner \etal \cite{turner01})
and one EPIC PN detector (Str\"uder \etal \cite{struder01}). 
Each MOS detector is sensitive in the 0.2--10\,keV band, 
while the PN detector is sensitive in the 0.1--15\,keV band.
In all observations, the three quoted detectors 
were operated in the {\it full window mode} 
($\sim30^{\prime}$ diameter field of view) 
with medium (observations 1, 2 and 3) 
and thin (observation 4) 
optical blocking filters.
We reduced EPIC data with the \X\ Science Analysis System
(SAS) version 5.4.1, by performing standard screening of
the EPIC data in order to exclude time intervals 
with high background levels and/or solar flares.
This resulted in shortened exposures of observations 1 and 3
as indicated in Table \ref{datatab}.
We applied the automatic source detection and analysis pipeline
summarised in the following section to all the four observations.
Observations 1, 3  and 4 
have high enough statistics to carry out a detailed 
timing analysis of more than half of the individual
point sources in them. Observation 2 has a somewhat lower exposure. 
Thus, section \ref{results} present only the results we obtained 
from observations 1, 3 and 4.
\begin{table}
\centering
\small
\caption{\X\ observations of M31 used in this analysis.\label{datatab}}
\small
\begin{tabular}{cccccc}
\hline
\hline
Observation & Date  & $T_{\rm start}$  & $T_{\rm stop}$ &  Exposure time MOS$^{a}$  & Exposure time PN$^{a}$ \\
     & (UT)  & (UT)             & (UT)           & (ks)                 & (ks)\\             
\hline
1 & 25/06/2000 & 08:12:41 & 20:59:33       & 34.8 (30.1)& 31.0 (26.5)\\
2 & 28/12/2000 & 00:01:33 & 03:43:23       & 12.2 (12.2)& 9.9 (9.9)\\
3 & 29/06/2001 & 06:15:17 & 22:20:20       & 54.9 (31.0)& 52.4 (31.0)\\
4 & 06/01/2002 & 18:00:56 & 11:52:53$^{b}$ & 63.0 (63.0)& 60.1 (60.1)\\
\hline
\end{tabular}
\begin{list}{}{}
\item[$^{a}$] Effective exposures used in our analysis. 
The number in parentheses is the exposure after exclusion of time
intervals with high background levels.
\item[$^{b}$] The following day.
\end{list}
\end{table}
In Fig. \ref{m31image} the central part of 
the MOS1 image of M31 extracted from the 2002 January 6 observation is shown.
%


\section{Search strategy}
\label{strategy}

For each observation listed in table \ref{datatab}
we filtered PN, MOS1 and MOS2 events in the energy band 0.2--15\,keV,
and merged the events of the two MOS detectors
into a single MOS1+MOS2 event list (task {\it merge}).
We then produced images of the PN and MOS1+MOS2 fields
binned in square pixels of $2^{\prime\prime}\times 2^{\prime\prime}$. 
Note that this is smaller than 
the FWHM of the point spread function in the center of the field 
for both PN and MOS, which is
$\sim 6^{\prime\prime}$ for the PN and 
$\sim 4^{\prime\prime}$ for the MOS (\X\ Users' Handbook V2.1 \cite{XMMhandbook}). 
We detected all sources above the $5\sigma$ threshold
in the two images (task {\it ewavelet}).
For a description of the wavelet based source detection algorithm used by the
task {\it ewavelet} see Damiani \etal \cite{damiani97}.
For every detected source this task yields the source position and extension,
corresponding to the standard deviation of an equivalent Gaussian luminosity profile, 
i.e. the radius containing $\sim$66\% of the source counts.
In the longest exposure observation (observation 4) we found
$\sim 350$ sources 
in the PN image and $\sim 180$ in the MOS image. 
We removed from the source lists
some spurious sources close to the image edge, 
CCD borders and bad pixel lines 
(this was important especially in PN images). 

After that we extracted light curves of all detected point sources
and applied the barycentric correction to the photon arrival times.
We used an extraction radius 
of two times the source extension calculated by {\it ewavelet}
(i.e. $2\sigma$ of the equivalent Gaussian luminosity profile; 
this contains about 98\% of the source counts) 
whenever there was no superposition with nearby sources, 
and an extraction radius of one time the source extension 
in all other cases.

Finally we considered all extracted light curves containing more than
200 photons (including background) 
in order to have enough statistics for timing analysis.
In this way, for instance, we reduced the total number of analysed 
sources to $\sim$140 (PN) and $\sim$90 (MOS) in observation 4. 
Detailed discussion on the complete set of sources
that we analysed will be presented in a forthcoming paper.

The light curves obtained were searched for pulsations by using
the power spectrum technique developed by Israel \& Stella (\cite{giallo96}).
In order to maximize signal detection sensitivity 
we calculated a single power spectrum with the highest Fourier resolution
$\Delta \nu = 1/\Delta T$
(with $\Delta T$ the time span covered by the observation).
In each power spectrum we searched for peaks above 
the detection threshold up to the Nyquist frequency 
$\nu_{\rm Nyq} = 1 /2 \Delta t$, 
with the binning time $\Delta t$ equal to the intrinsic time
resolution of the EPIC cameras 
(i.e. 73.4 ms for the PN and 2.6 s for the MOS).
The detection threshold 
for peaks in a single power spectrum 
containing $N_{trial}=\Delta T / 2 \Delta t$ frequencies
was set such that the probability of exceeding it by chance 
in a total number of trials equal to $N_{trial}$ times
the number $n$ of light curves analysed in the image was
$\epsilon=n*N_{trial}*\epsilon_{single\,trial}$,
with $\epsilon=2.326 \times 10^{-4}$
(corresponding to $3.5 \sigma$ in a Gaussian approximation).
Thus, the detection threshold for the analysis 
of a single power spectrum
was given by $\epsilon/n$.
In this way,
the minimum detectable amplitude (or pulsed fraction) 
for a sinusoidal signal 
(i.e. the sensitivity of the search to coherent periodic signals) 
for each source can be derived according 
to the prescription given by Vaughan \etal (\cite{vaughan94}).


\section{Results}
\label{results}

\subsection{Timing analysis}
\label{res-timingan}

The systematic search described in section \ref{strategy}
revealed both the 865 s pulsation of the
transient super-soft source XMMU\,J004319.4+411759 
(Osborne \etal \cite{osborne01})
visible in the 2000 June 25 observation (observation 1)
and quiescent in all other archival \X\ observations, 
and the 2.78\,hr dips in the 2002 January 6 observation (observation 4)
of  XMMU\,J004314.1+410724  in the globular cluster Bo\,158 
(Trudolyubov \etal \cite{trudo02}).

\begin{figure}
\centering
\includegraphics[angle=-90,width=12cm]{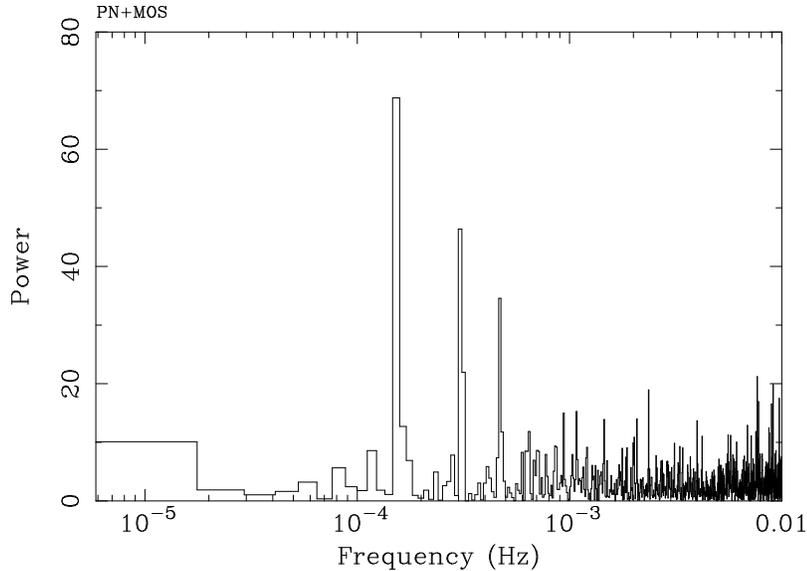}
\caption{Power spectrum of the combined PN and MOS light curves 
of the source \src\ 
corresponding to  the 2002 January 6 observation. 
The power spectrum is normalized according to Leahy \etal \cite{leahy83}.
\label{powspec}
}
\end{figure}

\begin{figure}
\centering
\includegraphics[angle=-90,width=10cm]{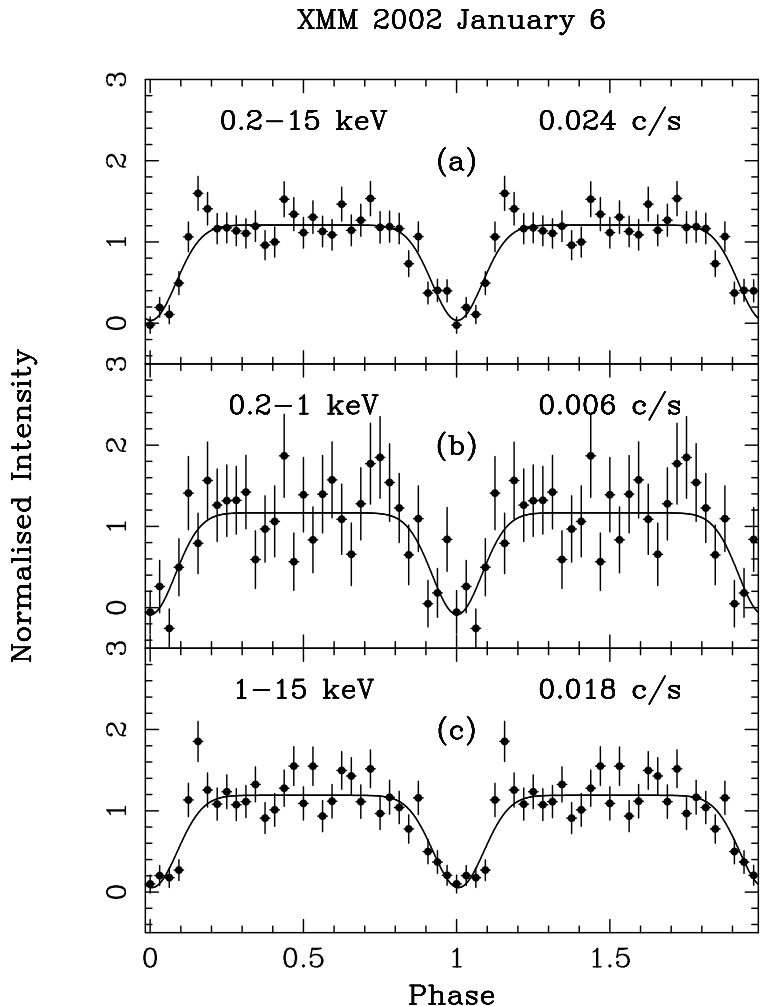}
\caption{Panel (a)  shows 
the combined PN+MOS light curve of \src\ 
from the \X\ observation of 2002 January 6 
(observation 4), in the energy band 0.2--15\,keV,
and folded at the best period of 6420\,s. 
The solid line represents the fit
of the data with the model described in the text 
(a constant minus a Gaussian). 
Panels (b) and (c) show the results obtained 
by folding at 6420\,s the combined PN+MOS light curve in the
0.2--1\,keV and in the 1--15\,keV energy band, respectively.
Each light curve is background subtracted and normalized 
to the average source count rate value 
given in the top right corner of the panel.
\label{foldedlc}
}
\end{figure}

We also found a significant peak 
at  $1.525879\times 10^{-2}$ Hz in the power spectrum of \src\ extracted from 
observation 4 MOS data. The source position we obtained,
RA = 00$^{\rm h}$43$^{\rm m}$08\fs611 
DEC = 41$^{\circ}$12\arcmin47\farcs58 (equinox 2000),
is to be considered accurate to better than 4$^{\prime\prime}$.
The power spectrum peak has a significance of $\sim 4.3 \sigma$ 
after taking into account the number of trial frequencies for this
source, as well as for the other sources in the MOS image of observation 4
for which the periodicity search could be carried out.
Visual inspection of the power spectra of the 2002 January 6 MOS and PN 
light curves of \src\ clearly reveals the presence
of the second and the third harmonics of the signal.
As a further confirmation, 
we calculated the power spectrum of the combined
PN and MOS light curves
(using the  XRONOS package, version 5.19) 
and obtained a significant increase
in the power of the signal and its harmonics. 
In Fig. \ref{powspec} the power spectrum of the PN+MOS light curve
of \src\ is shown. The signal fundamental plus 
two harmonics can be easily seen. 
An epoch folding search was carried out 
on the PN, MOS and PN+MOS light curves of \src\ 
from 2002 January 6 observation
aiming at a more accurate determination of the period value.
Following the prescription of Leahy (\cite{leahy87})
we obtained an average best period of $6420\pm80$\,s 
(error estimate at 67\% confidence level). 
For the background count rate estimation we used an annular region around
the extraction circle of \src\ (see also section \ref{res-spectralan}).
The background subtracted PN+MOS light curve folded
at a period of 6420\,s is shown in Fig. \ref{foldedlc} (panel a). 
The reference epoch $T_0$ corresponding to the minimum of the modulation 
is given in Table \ref{refepoch}, 
together with the source count rate. 
The minima are consistent with a zero count rate.
The dips last $\sim$25\% of each cycle and their fall and rise are
fairly continuous. A fit of the folded light curve in Fig.\ref{foldedlc}
with a simple model consisting of 
a constant and a Gaussian with negative normalization centered at the
minimum gives a nearly 100\%
amplitude of the dip with respect to the out-of-dip intensity  
and a $\sigma = (8.3 \pm 1.2) \times 10^{-2}$ (phase units).
See Table \ref{refepoch}.
The same results are 
obtained from the combined PN+MOS light curve
restricted to the 0.2--1\,keV and 1--15\,keV energy bands and folded
at the 6420\,s period (see Fig. \ref{foldedlc}, panels b and c).
The fit to the folded light curve gives 
a dip amplitude of 104\%$\pm$27\% and 95\%$\pm$12\%  
in the 0.2--1\,keV and in the 1--15\,keV energy band respectively, fully
compatible with the  98\%$\pm$11\% 
amplitude obtained in the 0.2--15\,keV
energy band.
We thus conclude that the data provide no evidence for an energy dependence 
of the modulation.

\begin{figure}
\centering
\includegraphics[angle=-90,width=10cm]{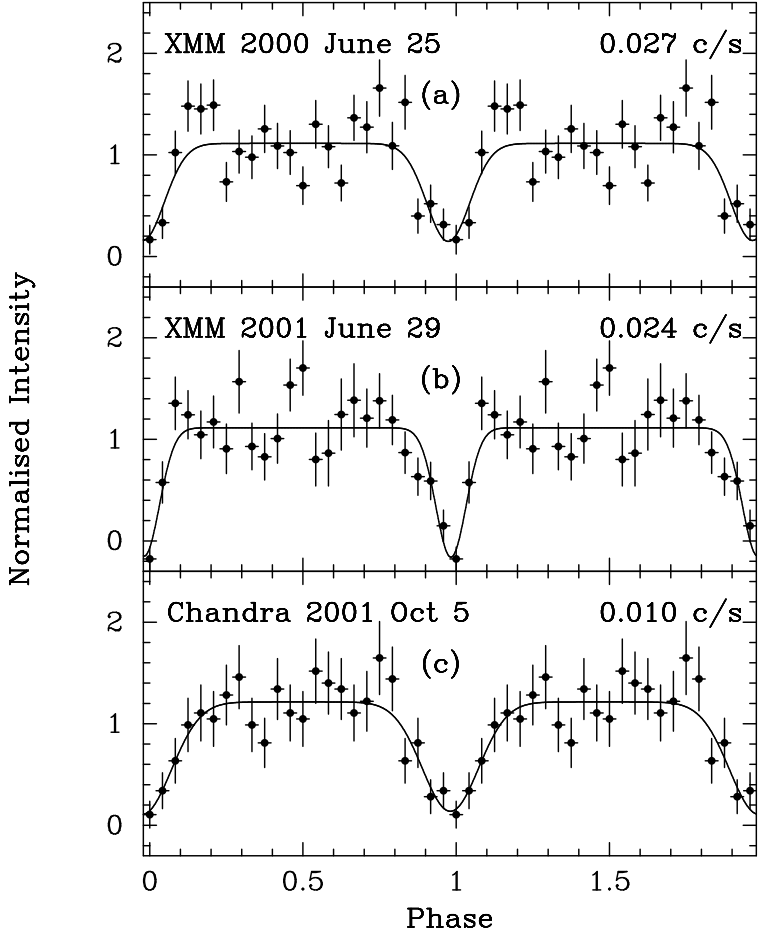}
\caption{We show the combined PN+MOS light curves of \src\ 
in the band 0.2--15\,keV folded at 107\,min period
extracted from \X\ observations of 2000 June 25 (a) and 2001 June 29 (b). 
The fit with a simple model consisting of a constant and
a Gaussian with negative normalization is also shown (solid line).
In panel (c) we show the folded light curve of an ACIS--S \C\
observation of the source of 2001 October 5.
Each light curve is background subtracted and
normalized to the source average count rate. This is given
in the top right corner of each panel.
\label{otherfoldedlc}}
\end{figure}

We searched for a modulation at a period around 6420\,s 
also in all the other \X\ observations of \src, 
using combined PN+MOS light curves of the source in all cases
in order to improve statistics. 
An epoch folding search yields
a best period of  6400$\pm$440\,s in observation 1 
and 6425$\pm$200\,s in observation 3. 
The errors are a factor 3--5 larger than in observation 4
and the best period values are consistent with one another.
Therefore we are justified in adopting the same period of 6420\,s 
to fold all the light curves.
We folded the \src\ light curves 
extracted from \X\ observations 1 and 3 
at the 6420\,s period and fitted the folded light curve with the
same model  used for observation 4
(a constant and a Gaussian with negative normalization).
We confirmed the presence of the signal at the same period
and with a modulation amplitude compatible with that previously found.  
Results are shown in Fig. \ref{otherfoldedlc} (panels a and b)
and summarised in Table \ref{refepoch}.

\begin{figure}
\centering
\includegraphics[angle=-90,width=10cm]{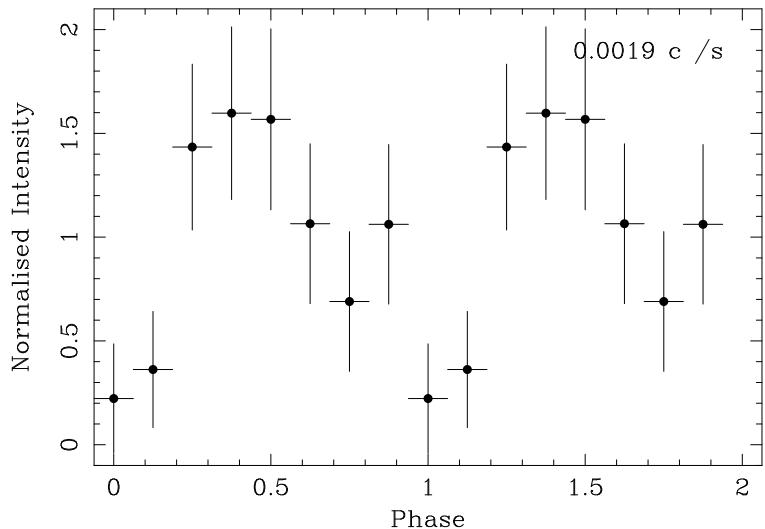}
\caption{Background subtracted and folded light curve of \src\ 
from the \C\ HRC--I observation of 2001 November 1.
The average source count rate, is given in top right corner of the panel.
\label{foldedHRClc}}
\end{figure}

In order to further investigate \src\ we used 
archival \C\ and \R\ observations of M31.
We selected the archival \C\ observations with the longest exposure, 
i.e. that  carried out  
on 2001 October 5  with the Imaging Spectrometer ACIS--S 
(Garmire \etal \cite{garmire92}; Bautz \cite{bautz98} and references therein)
for an effective exposure of $\sim$37.7 ks and the
2001 November 1 observation with the HRC--I 
in imaging mode for an effective exposure of $\sim$46.8 ks 
(Murray \etal \cite{murray97}).
We used standard event processing and filtering procedures 
in the CIAO package, Version 2.2.1,
and extracted light curves of \src\ from circular regions of diameter about   
twice the FWHM of the local Point Spread Function (PSF), 
in the 0.3--10\,keV band. 
We folded the \src\ light curves obtained from \C\ data at the 6420\,s period.
The dips are clearly detected in the ACIS--S \C\ 
observation carried out on 2001 October 5 
(see panel c in Fig. \ref{otherfoldedlc}) and their properties are 
consistent with those of the \X\ observations (see Table \ref{refepoch}).
The best period obtained with an epoch folding search 
on the ACIS--S \C\  observation of 2001 October 5
is 6437$\pm$200\,s, consistent with the period during observation 4.
There is marginal evidence for the 107\,min signal also in 
HRC--I \C\ observation
of 2001 November 1 (see Fig. \ref{foldedHRClc}) 
though the epoch folding search do not show any significant peak. 

We then considered the longest exposure \R\ archival observations of M31 
containing \src\  within ten arc--minutes from the center 
of the field of view. 
These are the 1996 January 7 HRI observation (85.5 ks exposure) 
and the 1991 July 27 PSPC observation (30 ks exposure). 
The \R\ light curves of \src\ obtained contain too few photons 
($\sim 60$ and $\sim 80$ net source counts respectively) 
to confirm the presence of the dips.

\begin{table}
\centering
\small
\caption{Parameters of the observed modulation \label{refepoch}}
\small
\begin{tabular}{ccccccccc}
\hline
\hline
& Mission & Date (UT) & $T_{0}$\,$^{a}$ (MJD) & \multicolumn{2}{c}{SOURCE\,$^{b}$ (counts/s)} & $\sigma$\,$^{c}$ & A\,$^{d}$ \\ 
&         &           &                       & PN             &   MOS                        &                  &           \\
\hline
1 & XMM     & 25/06/2000 & 51720.5201$\pm 1.5\times10^{-3}$ & 1.7$\pm$0.1 & 0.98$\pm$0.08 
& 0.07$\pm$0.015 & 88\%$\pm$20\%\\ 
3 & XMM     & 29/06/2001 & 52089.4458$\pm 7\times 10^{-4} $ & 1.4$\pm$0.2 & 1.0$\pm$0.1 
& 0.05$\pm$0.015 & 115\%$\pm$18\%\\
4 & XMM     & 06/01/2002 & 52280.7957$\pm 9\times 10^{-4} $ & 1.32$\pm$0.09 & 1.08$\pm$0.06
& 0.08$\pm$0.01  & 98\%$\pm$11\%\\
\hline
\hline
& Mission & Date (UT)  & $T_{0}$\,$^{a}$ (MJD)           & \multicolumn{2}{c}{SOURCE\,$^{b}$ (counts/s) } & $\sigma$\,$^{c}$ & A\,$^{d}$ \\ 
&         &            &                                 & \multicolumn{2}{c}{ACIS}                      &                  &           \\
\hline
& Chandra & 05/10/2001 & 52187.057$\pm 2 \times10^{-3} $ & \multicolumn{2}{c}{0.99$\pm$0.05} 
& 0.07$\pm$0.02  & 92\%$\pm$17\%\\
\hline
\end{tabular}
\begin{list}{}{}
\item[$^{a}$] Reference epoch corresponding to the minimum of the modulation
\item[$^{b}$] Source count rate in units of $10^{-2}$ counts/s 
              in the 0.2--15\,keV band for the \X\ observations and in the
              0.3--10\,keV band for the \C\ observation. 
              The MOS column refers to the MOS1+MOS2 data.
\item[$^{c}$] Standard deviation of the Gaussian dip in phase units. 
              Errors are at the 90\% confidence level.
\item[$^{d}$] Amplitude of the modulation with respect to the out-of-dip 
              intensity. 
              Errors are at the 90\% confidence level.
\end{list}
\end{table}

Note that the four folded light curves of \src\ 
obtained from the \X\ and \C\ data
can be reasonably well fitted also by a square--wave.
For this we adopted a model of the form
$Y(F)=D\left[\arctan(A(F_1-F))/\pi+ \arctan(A(F-F_2))/\pi\right]+C$, 
where $A$, $F_1$, $F_2$, $D$ and $C$ are the parameters of the model 
and $F$ represent the phase. 
For very large values of $A(F_2-F_1)$ ($>1000$) this expression
produces an eclipse-type behaviour with sharp ingress and egress
at phases $F_1$ and $F_2$, respectively. When $A(F_2-F_1)$  is smaller 
the ingress and egress become progressively smoother, $F_1$ and $F_2$ being
the half depth phases. When $A(F_2-F_1)\simlt 100$ the profile turns into a 
dip--like one with no flat minimum phase interval.

\begin{figure}
\centering
\includegraphics[angle=-90,width=20cm]{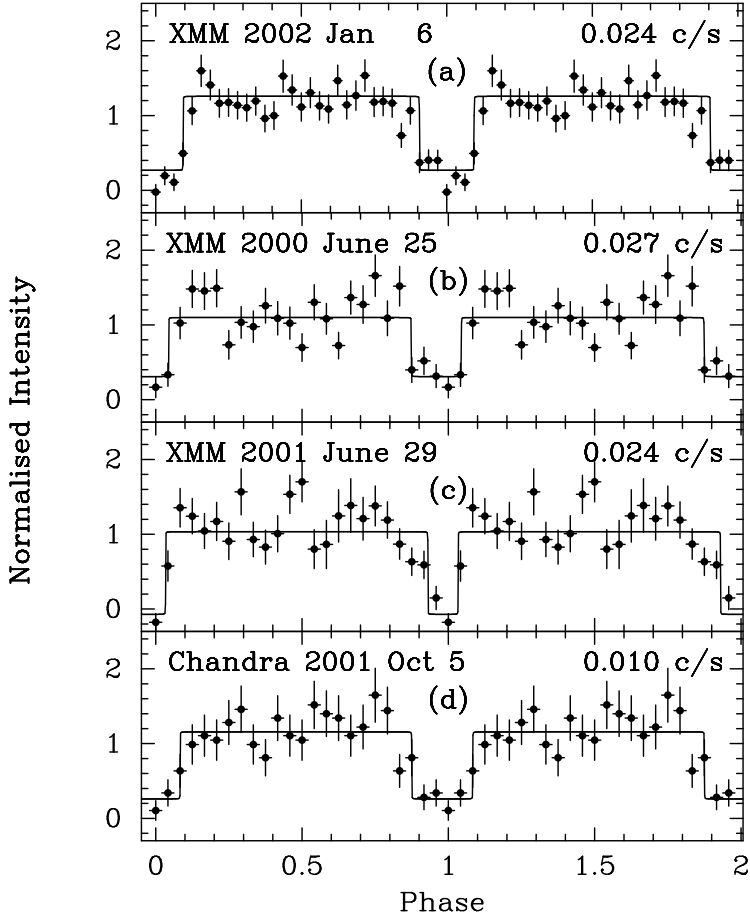}
\caption{
Square--wave fit
of the background subtracted and folded light curves of \src\ 
from the \X\ and \C\ observations showing the 107\,min pulsation. 
The  satellite, observation date and average source count rate 
are given in each panel.
The data are the same as in Fig. \ref{foldedlc} 
panel (a) and in Fig. \ref{otherfoldedlc} 
panels (a), (b) and (c).
\label{foldedsqwa}}
\end{figure}

We fit all our folded light curves to the square--wave model
with $A$ frozen to $10^4$,
the $\chi^2$ being insensitive to changes in $A$ within one
order of magnitude from this value. 
The results are shown in Fig. \ref{foldedsqwa}.
The square--wave model fit provides a reduced $\chi^2$ slightly
closer to 1 than the Gaussian fit, but uncertainties in the parameters
D and C (needed for the eclipse depth estimation) are large.
According the square--wave model fit 
the \X\ 2002 January 6 light curve profile  
shows an eclipse lasting 0.19$\pm$0.03 phase units
with a 92\%$\pm$38\% decrease of the source flux during the eclipse
(90\% confidence level errors).
In the \X\ 2000 June 25 light curve the eclipse lasts 0.21$\pm$0.06
phase units and produces a 84\%$\pm$45\% flux decrease; in the
\X\ 2001 June 29 it lasts 0.125$\pm$0.005 and gives
a 104\%$\pm$60\% flux decrease and finally, in the \C\ 2001 October 5
observation the eclipse lasts 0.21$\pm$0.03 and
gives an 89\%$\pm$49\% flux decrease. 

Because of these results we conclude that poor 
statistics does not allow us to distinguish between a smooth
profile of the folded light curve profile (as the one represented by
the Gaussian model) or a sharp eclipse--like light curve 
profile.

\subsection{Spectral analysis}
\label{res-spectralan}

We extracted phase-averaged spectra of \src\
in the 0.2--15\,keV energy band from the PN and MOS data
of all the \X\ observations in which 
the 107\,min modulation was detected (observations 1, 3 and 4). 
We generated the relevant spectral response matrices 
(SAS tasks {\it rmfgen} and {\it arfgen}). 
The background spectrum 
was extracted from the same annular region around 
the source extraction circle 
used for the background count rate estimation.
These spectra were analysed simultaneously by fitting 
the following spectral models: 
absorbed power law, absorbed power law and blackbody, absorbed power law
with exponential cut off, absorbed thermal bremsstrahlung, 
absorbed Comptonization model (Titarchuk \cite{tit94}; Titarchuk \& Lyubarskij \cite{tit95})
with spherical or disk geometry.
Given the relatively poor statistics of our spectra,
it is perhaps not surprising that a simple absorbed power law provides
a statistically acceptable fit (see below). 
Nonetheless we fit also a larger sample of two components spectral models
that are often used to describe the spectra of various classes of 
Galactic pulsating X--ray sources in order to try to draw 
some closer analogies.
The spectral fitting was performed using the same model parameters 
for the PN, MOS1 and MOS2 phase-averaged spectra. 
We allowed for independent normalizations of the three spectra 
in order to account for calibration differences between the EPIC cameras. 
Results from the spectral fitting are given in Table \ref{specpar}.
Only models yielding a reduced $\chi^2$ less than 2 are reported.
The model consisting of an absorbed power law with an exponential cut off
was rejected because it yields a cutoff energy greater than 100\,keV,
i.e. a spectrum virtually indistinguishable 
from a power law in the \X\ energy band.
Note that the absorbed Comptonization model (Titarchuk \cite{tit94}) 
gives an acceptable fit
only by assuming a disk geometry for the Comptonising region.

\begin{figure}
\centering
\includegraphics[angle=-90,width=12cm]{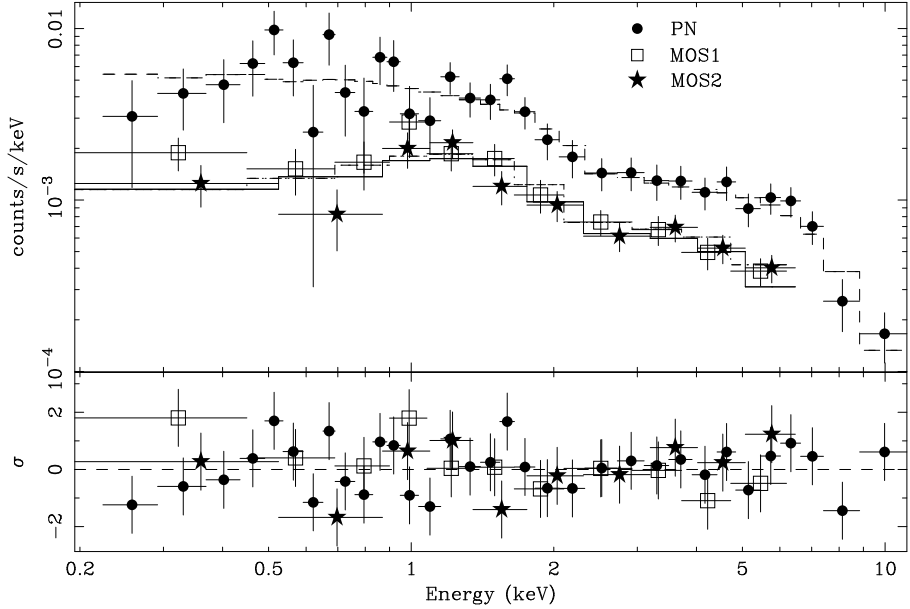}
\caption{phase-averaged spectra of \src\ extracted from 
PN, MOS1 and MOS2 data
of the \X\ observation carried out on 2002 January 6. 
The resulting simultaneous fit with the absorbed power law model is shown
(see table \ref{specpar}). 
The dashed line gives the best fit model for the PN spectrum, 
the dot-dashed is for the MOS1 and the solid line for the MOS2.
\label{pnm1m2sp} }
\end{figure}

The model consisting of a simple absorbed power law, 
gives the best reduced $\chi^2$ in all observations,  
but provides an upper limit to the absorption column 
at 95\% confidence level that is lower than the Galactic value of 
$7 \times 10^{20}$ cm$^{-2}$ 
derived from the HI distribution map by Dickey \& Lockman \cite{dilo90}.
The latter has a very limited angular resolution 
($1^{\circ}\times 1^{\circ}$). It was estimated 
that small scale structure in the HI distribution in the sky introduces
errors in the column density estimates of $\pm 1 \times 10^{20}$\,cm$^{-2}$ 
(90\% confidence level, see Elvis \etal \cite{elvis86}, Appendix B). 
The upper limits to $N_{\rm H}$ derived from the absorbed power-law fit 
are below the 90\% confidence interval for the Galactic value.
The fits with the Comptonization model has
the same problem  in two out of three observations.
The inclusion of a black body component in the absorbed power law model
(this is done in analogy with the X--ray spectra modeling used for several
Galactic dipping sources, see e.g. XB\,1916--053, Church \etal \cite{church97}; XB\,1254--690, 
Smale, Church \& Ba{\l}uci\'{n}ska--Church \cite{smale02}; 
X\,1624--49,  Church \& Ba{\l}uci\'{n}ska--Church \cite{churchbalu95}) 
gives plausible upper limits to the absorption column,
but is not required statistically since the power-law fit 
is already acceptable.
The phase-averaged spectra of \src\ obtained from the 2002 January 6 
observation are shown in Fig. \ref{pnm1m2sp}.

In order to investigate further the low--level absorption 
indicated by the above fits, 
we also fit groups of PN, MOS1 and MOS2
phase-averaged spectra with the three above-discussed models, while freezing 
$N_{\rm H}$ at the Galactic value (which is to be considered a lower limit
to the absorption for a source located in M31). 
We obtain reduced $\chi^2$ values of
1.2 (power law), 1.03 (power law plus blackbody) 
and 1.1 (comptonization) for observation 1,
of 1.4, 0.96 and 1.2 for observation 3
and of 1.3, 0.76 and 1.06 for observation 4 
respectively and parameter values consistent 
with the corresponding ones listed in Table \ref{specpar}
to within the errors. 
Note also that the upper limits to $N_{\rm H}$ increases 
considerably (up to values consistent with the Galactic value in some cases) 
when allowing for independent absorption columns 
in the PN, MOS1 and MOS2 spectra. 
The 2002 January 6 PN, MOS1 and MOS2 phase-averaged spectra fit 
in this way with the absorbed power--law model gives
a $\chi^2$ of 48.1 (47 d.o.f.), 
a photon index of $0.79_{-0.09}^{+0.1}$ and 
$N_{\rm H} < 4 \times 10^{20}$ cm$^{-2}$ for the PN spectrum, 
$N_{\rm H} < 6.8 \times 10^{20}$ cm$^{-2}$ for the MOS1 spectrum, 
$N_{\rm H} < 1.4 \times 10^{20}$ cm$^{-2}$ for the MOS2 spectrum.
Fitting again the three spectra 
with the absorbed Comptonization model (disk geometry), 
we obtain a $\chi^2$ of  37.20 (45 d.o.f.), 
an input photon temperature of $kT_0 < 0.12$\,keV,
Comptonising electron temperature of $kT_e = 3.3_{-0.8}^{+2.6}$\,keV,
and optical depth $\tau = 12.5_{-1.75}^{+1.8}$ and
$N_{\rm H} < 6.8 \times 10^{20}$ cm$^{-2}$ for the PN spectrum, 
$N_{\rm H} < 11.5 \times 10^{20}$ cm$^{-2}$ for the MOS1 spectrum, 
$N_{\rm H} < 1.7 \times 10^{20}$ cm$^{-2}$ for the MOS2 spectrum.
Based on this analysis we conclude that the low 95\%  upper 
limits to $N_{\rm H}$ obtained may be related to calibration 
uncertainties of the EPIC cameras. 
It is worth noting that  \src\ is not the only source we found showing
this anomalous behaviour. Also the bright source XMMU\,J004232.0+411314 
show spectra with typical photon index $\sim 1$ and 
$N_{\rm H} < 0.6 \times 10^{20}$ cm$^2$ at 95\%  confidence level
in the 2002 January 6 observation. In this case the upper limit to $N_{\rm H}$
grows to $0.9 \times 10^{20}$ cm$^2$ and to $1.2 \times 10^{20}$ 
in MOS1 and MOS2 spectra respectively when allowing for independent
$N_{\rm H}$ parameters in the fit (we did not extract PN spectra 
because in the PN image the source is more than 50\% covered 
by CCD edges and hot pixel lines). XMMU\,J004232.0+411314 corresponds
to source r2-32 in the ACIS--I \C\ catalogue by Kong \etal (\cite{kong02}). 
In the quoted paper (Table 8) the source was listed among the 20 
brightest sources in M31 and its spectrum was well represented by  
a power--law spectral fit in the 0.5--10\,keV band
with photon index $1.3_{-0.1}^{+0.2}$ and 
$N_{\rm H} < 5 \times 10^{20}$ cm$^2$ at the 95\%  confidence level, 
i.e. an upper limit to the absorption column lower than the Galactic value.
Based on the absorbed flux values in the 0.3--10\,keV band given 
in Table \ref{specpar}, 
we can conclude that the luminosity of \src\ was roughly constant 
over the $\sim1.5$\,yr time span covered by \X\ observations, at a level of
$\sim 1\times10^{37} \left(d / d_{\rm M31} \right)^{2}$\,erg\,s$^{-1}$, 
where $d$ is the actual distance of \src\ from us and 
$d_{\rm M31}$ the M31 distance.
\begin{table}
\centering
\small
\caption{Best--fit model parameters of energy spectra of \src\ 
$^{\rm a}$ \label{specpar}}
\small
\begin{tabular}{ccccccc}
\hline
\hline
Parameter                      & \multicolumn{6}{c}{Observation Date}\\
                               & \multicolumn{2}{c}{2000 June 25}    & \multicolumn{2}{c}{2001 June 29}   & \multicolumn{2}{c}{2002 January 6} \\ 
                               & \multicolumn{2}{c}{(observation 1)} & \multicolumn{2}{c}{(observation 3)}& \multicolumn{2}{c}{(observation 4)} \\ 
\hline
\multicolumn{7}{c}{Absorbed Power Law [wabs*powerlaw]}\\
\hline
N$_{\rm H}^{\;\;\;\; \rm b}$   &$\le 0.058$              &&$\le 0.023 $           &&$\le 0.012$ &\\
Photon Index                   &$0.84_{-0.12}^{+0.17}$   &&$0.75_{-0.13}^{+0.13}$ &&$0.78^{+0.09}_{-0.08}$&\\
Flux$^{\;\; \rm c}$            & $1.71_{-0.10}^{+0.13}$  &$\left(1.76_{-0.10}^{+0.13}\right)$
                               & $1.56_{-0.12}^{+0.13}$  &$\left(1.56_{-0.12}^{+0.13}\right)$ 
                               & $1.68_{-0.07}^{+0.07}$  &$\left(1.68_{-0.07}^{+0.07}\right)$\\ 
$\chi^{2}$(d.o.f)              &$21.53(20)$              &&$27.48(25)$            &&$48.13(49)$&\\
\hline
\multicolumn{7}{c}{Absorbed Power Law plus Black Body [wabs*(powerlaw+blackbody)]}\\
\hline
N$_{\rm H}^{\;\;\;\; \rm b}$     &$\le 0.32$             &&$0.07^{\;\;\rm h}$     &&$\le 0.098$&\\
Photon Index                     &$1.81_{-1.5}^{+0.9}$   &&$2.14_{-0.6}^{+0.3}$   &&$1.73^{+0.82}_{-0.7}$&\\  
kT$_{\rm bb}^{\;\;\;\; \rm d}$   &$2.3_{-0.54}^{+2.7}$   &&$2.07_{-0.4}^{+0.96}$  &&$2.55^{+1.4}_{-0.5}$&\\  
Flux$^{\;\; \rm c}$              & $1.69_{-0.35}^{+0.85}$ &$\left(1.81_{-0.37}^{+0.90}\right)$
                                 & $1.50_{-0.24}^{+0.47}$ &$\left(1.58_{-0.25}^{+0.49}\right)$ 
                                 & $1.73_{-0.23}^{+0.37}$ &$\left(1.79_{-0.23}^{+0.37}\right)$\\
Flux (black body)$^{\;\; \rm c}$ &$1.14_{-0.29}^{+0.63}$  &$\left(1.15_{-0.29}^{+0.64}\right)$
                                 & $1.20_{-0.19}^{+0.41}$ &$\left(1.20_{-0.19}^{+0.42}\right)$   
                                 & $1.21_{-0.19}^{+0.32}$ &$\left(1.21_{-0.19}^{+0.32}\right)$\\
Flux  (power law)$^{\;\; \rm c}$ &$0.55_{-0.07}^{+0.22}$  &$\left(0.66_{-0.08}^{+0.26}\right)$
                                 & $0.30_{-0.07}^{+0.06}$ &$\left(0.36_{-0.06}^{+0.07}\right)$ 
                                 & $0.52_{-0.04}^{+0.04}$ &$\left(0.58_{-0.05}^{+0.05}\right)$\\
$\chi^{2}$(d.o.f)                &$17.25(16)$            &&$21.09(22)$            &&$34.77(45)$&\\
\hline
\multicolumn{7}{c}{Absorbed Comptonization Model [wabs*comptt]}\\
\hline
N$_{\rm H}^{\;\;\;\; \rm b}$   &$\le 0.096$              &&$\le 0.044$            &&$\le 0.037$&\\
kT$_{0}^{\;\;\;\; \rm e}$      &$\le 0.23$               &&$\le 0.17$             &&$\le 0.15$&\\
kT$_{\rm e}^{\;\;\;\; \rm f}$  &$3.2_{-1.0}^{+49.2}$     &&$2.84_{-0.85}^{+22.7}$ &&$3.2_{-0.8}^{+2.2}$&\\
$\tau^{\;\;\;\; \rm g}$        &$11.5_{-3.25}^{+3.5}$    &&$14.1_{-3.0}^{+3.1}$   &&$13_{-1.1}^{+2.0}$&\\
Flux$^{\;\; \rm c}$            &$1.73_{-0.45}^{+0.72}$   &$\left(1.77_{-0.46}^{+0.74}\right)$ 
                               &$1.60_{-0.21}^{+0.77}$   &$\left(1.60_{-0.21}^{+0.77}\right)$  
                               &$1.77_{-0.23}^{+0.56}$   &$\left(1.78_{-0.23}^{+0.57}\right)$\\
$\chi^{2}$(d.o.f)              &$20.20(18)$              &&$22.81(23)$            &&$40.6(47)$&\\ 
\hline
\end{tabular}

\begin{list}{}{}
\item[$^{\rm a}$] Results of simultaneous fit of EPIC--PN,  MOS1 and MOS2 
phase-averaged spectra in the 0.2--15\,keV energy band.
Parameter errors correspond to 90\% confidence level.
\item[$^{\rm b}$] Upper limit at 95\% confidence level 
to the equivalent absorbing hydrogen column density in units 
of $10^{22}$ cm$^{-2}$
\item[$^{\rm c}$] Observed flux in the 0.3--10\,keV energy range in units of 
$10^{-13}$\,erg\,s$^{-1}$ cm$^{-2}$. 
Average value from PN, MOS1 and MOS2. 
The value of the unabsorbed flux is in parentheses.
Errors are calculated from the uncertainty in the normalization at the 
67\% confidence level.
\item[$^{\rm d}$] Blackbody temperature in keV
\item[$^{\rm e}$] Upper limit to the temperature of soft photons in keV
\item[$^{\rm f}$] Electron temperature in keV
\item[$^{\rm g}$] Thomson optical depth for the disk geometry
\item[$^{\rm h}$] Parameter frozen to the Galactic value
\end{list}
\end{table}

\begin{figure}
\centering
\includegraphics[angle=-90,width=12cm]{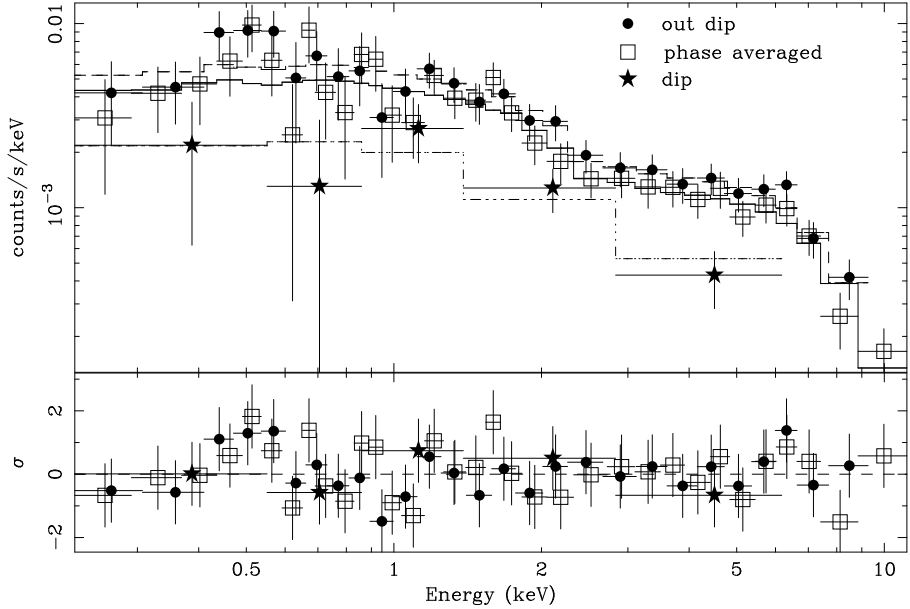}
\caption{out-of-dip, dip and phase-averaged spectra of \src\ 
extracted from PN data of the \X\ observation carried out on 2002 January 6. 
The resulting simultaneous best fit with the absorbed power law model
with photon index $\Gamma=0.78$
is shown (see Tables \ref{specpar} and \ref{specpar2}). 
The dashed line gives the best fit model for the out-of-dip spectrum, 
the dot-dashed is for the dip spectrum and the solid line 
for the phase-averaged spectrum.
\label{outinsp} }
\end{figure}

In order to investigate changes in the energy spectrum during the dips, 
we  divided the PN, MOS1 and MOS2
events into two phase intervals 
corresponding to the out-of-dip 
(from phase 0.1 to phase 0.85)
and dip intensity (from phase 0.85 to 0.1).
We extracted spectra from events in the two quoted phase intervals,
thus obtaining  out-of-dip and dip spectra.
The MOS1 and MOS2 dip spectra of observations 1 and 3 had less
than 30 counts each (including background) 
and therefore were not used in any spectral fit.
For each observation we fit the PN out-of-dip and dip spectra
simultaneously with the phase-averaged spectrum, 
using two different models: 
an absorbed power law and an absorbed Comptonization model. 
Independent normalizations and absorption columns for the 
three spectra were used, while the other parameters were 
frozen to the best fit values in Table \ref{specpar}.
To within the obvious statistical limitations this approach
is justified by the fact that 
simultaneous fits of the PN, MOS1 and MOS2 out-of-dip spectra alone
give results consistent with those obtained for the phase-averaged spectra
listed in Table \ref{specpar}. A similar check cannot be done with dip spectra
alone because of poor statistics, and the usage of the same spectral model 
fitting out-of-dip and phase-averaged spectra represents 
the simplest assumption.
Results are shown in Table \ref{specpar2}. 
Note that for the 2002 January 6 observation 
also the MOS1 and MOS2 phase-averaged, out-of-dip and dip spectra 
were fit simultaneously to the PN data.
In this case we used nine independent normalizations 
for the nine spectra and  three independent $N_{\rm H}$ parameters, 
one for the three (PN, MOS1 ans MOS2)  phase-averaged spectra, 
one for the out-of-dip spectra and one for the dip spectra. 
The PN dip and out-of-dip spectra 
from the 2002 January 6 observation are shown in fig. \ref{outinsp}
together with their best fit power law model. 
The resulting 95\%  confidence upper limit 
on the average increase in the (neutral) 
absorbing column density $N_{\rm H}$ during the dips 
was $9.7 \times 10^{22}$ cm$^{-2}$ in observation 1, 
$9 \times 10^{21}$ cm$^{-2}$ in observation 3  
and  
$2 \times 10^{21}$ cm$^{-2}$ in observation 4. 
(The tighter limit obtained for observation 4 clearly reflects the 
longer exposure time in that observation).
However the above limits should be taken as indicative, since 
in order to increase the S/N of the dip spectrum was integrated over 
the whole duration of the dips (as opposed to a shorter phase interval
around the dip minimum). 
\begin{table}
\centering
\small
\caption{Best--fit model parameters of energy spectra of \src\ 
$^{\rm a}$ \label{specpar2}}
\small
\begin{tabular}{ccccccc}
\hline
\hline
Parameter                                & \multicolumn{6}{c}{Observation Date}\\
                                         & \multicolumn{2}{c}{2000 June 25 }         & \multicolumn{2}{c}{2001 June 29  }       &\multicolumn{2}{c}{ 2002 January 6  }     \\
                                         & \multicolumn{2}{c}{(observation 1)   }    & \multicolumn{2}{c}{(observation 3)    }  &\multicolumn{2}{c}{ (observation 4) }     \\ 
                                         &  \multicolumn{2}{c}{  PN data     }       &  \multicolumn{2}{c}{   PN data   }       &\multicolumn{2}{c}{ PN,MOS1 and MOS2 data   }           \\ 
\hline
\multicolumn{7}{c}{Absorbed Power Law [wabs*powerlaw]}\\
\hline
N$_{\rm H}$ (out-of-dip)$^{\;\;\;\; \rm b}$ &$\le 0.023$   &         &$\le 0.037$   &        &$\le 0.01$   &  \\ 
N$_{\rm H}$ (    dip)$^{\;\;\;\; \rm b}$ &$\le 9.7$     &         &$\le 0.93$    &        &$\le 0.22$   &  \\ 
Flux (out-of-dip)$^{\;\;\;\; \rm c}$        &$2.19_{-0.13}^{+0.14}$&$\left(2.19_{-0.13}^{+0.15}\right)$
                                         &$1.89_{-0.14}^{+0.14}$&$\left(1.89_{-0.14}^{+0.14}\right)$        
                                         &$2.02_{-0.07}^{+0.07}$&$\left(2.02_{-0.07}^{+0.07}\right)$\\ 
Flux (    dip)$^{\;\;\;\; \rm c}$        &$1.01_{-0.31}^{+0.87}$&$\left(1.09_{-0.33}^{+0.94}\right)$
                                         &$1.09_{-0.28}^{+0.31}$&$\left(1.09_{-0.28}^{+0.31}\right)$        
                                         &$0.69_{-0.13}^{+0.16}$&$\left(0.69_{-0.13}^{+0.16}\right)$\\ 
$\chi^{2}$(d.o.f)                        &$39.74(26)$   &         &$27.48(24)$   &        &$85.85(95)$  &  \\ 
\hline
\multicolumn{7}{c}{Absorbed Comptonization Model [wabs*comptt]}\\
\hline
N$_{\rm H}$ (out-of-dip)$^{\;\;\;\; \rm b}$ &$\le 0.044$    &        &$\le 0.048$   &        &$\le 0.03$     & \\ 
N$_{\rm H}$ (    dip)$^{\;\;\;\; \rm b}$ &$\le 32$       &        &$\le 0.90$    &        &$\le 0.19$     & \\ 
Flux (out-of-dip)$^{\;\;\;\; \rm c}$        &$2.14_{-0.16}^{+0.44}$ &$\left(2.14_{-0.16}^{+0.44}\right)$        
                                         &$1.93_{-0.18}^{+0.53}$ &$\left(1.93_{-0.18}^{+0.53}\right)$        
                                         &$2.11_{-0.12}^{+0.34}$ &$\left(2.13_{-0.12}^{+0.34}\right)$\\ 
Flux (    dip)$^{\;\;\;\; \rm c}$        &$1.03_{-0.32}^{+0.84}$ &$\left(1.16_{-0.35}^{+0.91}\right)$        
                                         &$1.11_{-0.29}^{+1.22}$ &$\left(1.11_{-0.29}^{+1.22}\right)$ 
                                         &$0.73_{-0.15}^{+0.21}$ &$\left(0.73_{-0.15}^{+0.21}\right)$\\ 
$\chi^{2}$(d.o.f)                        &$33.12(25)$    &       &$22.81(23)$   &        &$70.79(94)$    & \\ 
\hline
\end{tabular}
\begin{list}{}{}
\item[$^{\rm a}$] Results of simultaneous fit of the 
out-of-dip,  dip and phase-averaged spectra of source \src.
Spectral parameters are frozen to the 
values indicated in Table \ref{specpar}.
\item[$^{\rm b}$] Upper limit at 95\% confidence level 
to the equivalent absorbing hydrogen column density in units of 
$10^{22}$ cm$^{-2}$
\item[$^{\rm c}$] Model observed flux in the 0.3--10\,keV 
energy range in units of $10^{-13}$\,erg\,s$^{-1}$ cm$^{-2}$.
The value of the unabsorbed flux is in parentheses.
Errors are calculated from the uncertainty in the normalization at the 
67\% confidence level.
\end{list}
\end{table}

\subsection{Other Observations}
\label{res-otherobs}

An X--ray source with a position consistent with that of
\src\ was detected with 
the \R--{\it HRI} (source [PFJ93]74, Primini, Forman \& Jones \cite{prifojo93}), 
\R--{\it PSPC} 
(sources 2RXP\,J004308.5+411251 and 
         2RXP\,J004308.3+411249 in the Second ROSAT PSPC Catalogue; 
source     RX\,J0043.1+4112 in the M31 second ROSAT PSPC survey Supper \etal \cite{supper01}), 
and  \C\ (source
CXOM31\,J004308.4+411247 in Kaaret \cite{kaaret02}, source
CXOM31\,J004308.6+411248 in Kong \etal \cite{kong02}, source
CXOM31\,J004308.6+411250 in Williams \etal \cite{williams03}). 

The observed flux of \src\ from the latter experiments
is roughly constant over a 10 years time span.
>From the published values of the source luminosity
we derive values of the integrated unabsorbed flux in the 0.3--10\,keV band of 
$\sim 3.5 \times 10^{-14}$ \,erg\,s$^{-1}$\,cm$^{-2}$  
from \R--{\it HRI} 1990 July 25 observation 
(Primini, Forman \& Jones \cite{prifojo93}),
$\sim 1.0 \times 10^{-13}$ \,erg\,s$^{-1}$\,cm$^{-2}$  
from \C--ACIS 1999--2000 observations 
(Kong \etal \cite{kong02}),
$\sim 8.5 \times 10^{-14}$ \,erg\,s$^{-1}$\,cm$^{-2}$  
from \C--HRC 1999--2001 observations 
(Williams \etal \cite{williams03}),
$\sim 3.4 \times 10^{-14}$ \,erg\,s$^{-1}$\,cm$^{-2}$  
from \C--HRC 2001 October 31 observation 
(Kaaret \cite{kaaret02}).
Note that both Primini, Forman \& Jones (\cite{prifojo93}) and 
Kaaret (\cite{kaaret02}) 
use a thermal bremsstrahlung distribution to fit the
spectra, with temperature $kT = 5$\,keV and 2\,keV 
and absorption column $N_{\rm H} = 7 \times 10^{20}$\,cm$^{-2}$ and 
$6.66 \times 10^{20}$\,cm$^{-2}$ respectively,
while Kong \etal (\cite{kong02}) and Williams \etal (\cite{williams03}) 
use a power law with photon index $\Gamma=1.7$ and 
absorption column $N_{\rm H} = 10^{21}$\,cm$^{-2}$.
These values are within a factor five from the observed flux
of $\sim 1.7 \times 10^{-13}$\,erg\,s$^{-1}$\,cm$^{-2}$ 
we obtain from \X\ observations in the 0.3--10\,keV band 
(see Table \ref{specpar} and section \ref{res-spectralan}).
This indicates that \src\ is a fairly stable X--ray source, 
and it is not an X--ray transient.
%


\section{Discussion}
\label{discussion}

The unprecedent throughput of \X\ 
led to the detection of a 107\,min modulation
from the X--ray source \src\ in the field of M31.
This source is very likely the \R--HRI source [PFJ93]74 
(Primini, Forman \& Jones \cite{prifojo93}).
The modulation is confirmed also 
by \C\ archival observations of the source.
The source flux was at a nearly steady level of 
$\sim$1.7$\times 10^{-13}$\,erg\,s$^{-1}$ cm$^{-2}$
in the 0.3--10\,keV band throughout all \X\ observations.
This value is within a factor five from
flux measurements obtained by other missions
spanning the previous decade. 
The flux above corresponds to a luminosity  of
$\sim 1\times10^{37} \left(d / 780~kpc \right)^{2}$\,erg\,s$^{-1}$, 
where $d$ is the distance of \src\ from us.
The source spectrum is well fit by an absorbed power law model with a 
photon index $\Gamma \sim 0.8$, but is also compatible with
an absorbed Comptonization model ($kT_{e} \sim 3$\,keV and $\tau \sim 13$).   
No other simple single-component spectral model succeeds in fitting the 
\X\ data.
Owing to poor statistics the $N_{\rm H}$ value could not be measured
from the \X\ spectra and only upper limits were derived.

The 107\,min modulation is due to a dip in the light curve
that covers about one fourth of the cycle
with an energy-independent intensity decrease 
compatible with 100\% at dip minimum.
The apparent smoothness of \src\ light curves might be due to poor
statistics.
An eclipse--like behaviour is also compatible with the folded
light curve of \src.
In this case, however, the ``eclipse'' 
duration would be of 0.12 phase units during the 
\X\ observation of 2001 June 29, 
inconsistent with the $\sim0.2$ duration 
obtained for the other three observations.
This indicates that the observed modulation
might not be compatible with a true eclipse by a companion star.

The possibility that \src\ is foreground source in the Galaxy cannot be 
ruled out at present. Indeed the somewhat lower value of $N_{\rm H}$, compared
to the galactic $N_{\rm H}$,  that we inferred from the analysis of the X--ray 
spectrum would appear to favor this possibility. 

\subsection{Galactic scenario}
\label{galacticscen}

For an assumed distance of 1\,kpc, roughly  corresponding to the 
thickness of the Galaxy in the direction of M31, the 
0.3--10\,keV luminosity of \src\ would be 
$\sim 2\times 10^{31}$\,erg\,s$^{-1}$, close to the low--end of
(but still compatible with) the quiescent X--ray luminosity of 
magnetic cataclysmic variables (CVs; 
Ramsay \etal \cite{ramsay94}, 
Ramsay \& Cropper \cite{ramcro03}).
The period of the X--ray modulation  
of \src\ is suggestive of the orbital period of
a polar or intermediate polar (IP). 
Orbital periods of polars and IPs are mostly concentrated 
in the in the 1--10 hr range (Downes \etal \cite{downes01}). 
Current limits on the magnitude of any optical counterpart to \src\ 
(V $>$ 21.7, R $>$ 20.2, see Kong \etal \cite{kong02}; 
Haiman \etal \cite{haiman94}) 
are also compatible with a magnetic CV interpretation 
(see Downes \etal \cite{downes01}). 
Dip-like X--ray modulations and/or on/off shaped eclipses 
reflecting the orbital motion of the binary are 
seen in both classes of polars and IPs. 
Examples of this are Fo Aqr, AO Psc 
(Hellier, Garlick \& Mason \cite{hellgarmas93})
and XY Arietis (showing eclipses, Hellier \cite{hellier97}) 
among IPs;
2A0311-227 (EF Eri, Beuermann, Thomas \& Pietsch \cite{beu91}),
DP Leo (showing eclipses, Schwope \etal \cite{schwope02})
and, to lesser extent, AN UMa (Ramsay \etal \cite{ramsay94}) 
among polars.
However, the longest duration of an eclipse
obtained through occultation of a compact accreting X--ray source
by a donor star filling its Roche-Lobe of radius $r_L$ 
in a binary system with orbital period $P$, binary separation $a$ 
and mass ratio $q = M_{donor}/M_X < 1$ can be calculated as
${1 \over P} 2 r_L / { 2 \pi a \over P } = { r_L \over \pi a}$
(in phase units).
Using the relation of  Eggleton (\cite{eggle83})
to express the dependence of the Roche-Lobe radius of the donor $r_L$
on $q$ we derive an upper limit to an eclipse duration of
$0.4/\pi \sim 0.13$ phase units.
A 0.2 phase units eclipse of the X--ray source cannot be due
to occultation by the companion star. 
Thus, in any interacting binary interpretation, 
the observed X--ray modulation of \src\ cannot be explained
as due to a true eclipse.

If the dip ingress/egress in the \src\ light curves 
sharp on the basis of the square--wave fit, 
a more plausible explanation for the observed modulation 
would be that of eclipses by the accretion column in a Polar.
In this case no limit to the eclipse duration would be expected
and occasional variations in the light curve profile may occur 
(see the case of EP Dra, Bridge \etal \cite{bridge03}). 

The alternative scenario for \src\ X--ray modulation 
at the orbital period would be occultation of an IP 
by the plasma bulge due to impact of the 
accretion stream on the accretion disk. 
Such a mechanism, at work both in IPs and low mass
X--ray binaries (LMXRBs),
might produce a relatively 
long dip which, when averaged, has often a fairly smooth profile,
consistent with the Gaussian fit to the folded light curves
that we described in section \ref{res-timingan}.
The apparent
energy--independence of the dips (to within the uncertainties)
might be accounted for if the plasma cloud producing  
the dips had a very high degree of ionization, 
and the flux reduction were due to Thomson scattering
rather than photo-electric absorption. 
In the IP scenario the ionization front due to the central X--ray source 
cannot reach the edge of the accretion disk:
to produce the $> 90$\% energy independent decrease of the source luminosity, 
an electron column $N_{e}$ 
of more than $3.5 \times 10^{24}$\,cm$^{-2}$ would be required.
In order to have this amount of ionized matter (with solar composition) 
in a cloud of thickness $\Delta R$ at a distance R from the central source 
emitting an X--ray luminosity $L_X$,  
the ionization parameter $\xi={ L_{X} \Delta R \over N_{e} R^2 }$ 
should be larger than 1000
(Tarter, Tucker \& Salpeter \cite{tartuck69}; 
Hatchett, Buff \& McCray \cite{hatchbuffmccray76}). 
We define $\epsilon={\Delta R \over R}$, i.e. the depth of the material 
that produces the dips
expressed as a fraction of its distance from the source. We conservatively 
assume $\epsilon = 1$.
Thus, in the case of \src\ the distance from the central X--ray source
of the material causing the dips should be
$R = {L_{X} \epsilon \over N_{e} \xi } <  10^{-3} {L_{X} \over N_{e}}$.
For an X--ray luminosity of $L_X \sim 2 \times 10^{31}$\,erg\,s$^{-1}$
this maximum distance amounts to less than 1\,km, i.e. orders of magnitude
lower than the radius of any bulge in the accretion stream/disk interaction.

Most IPs display soft X--ray modulation
at the white dwarf (WD) spin period 
(33 s to 2 h, but always shorter than the orbital period, 
Patterson \cite{patterson94}) 
with energy--dependent amplitudes in the 10-40\% range.
No evidence was found for such a distinct X--ray modulation
(possibly, in the soft X--ray range) resulting from the white dwarf rotation 
as observed in a number of IPs, in the case of \src.
We note, however, that X--ray pulsations with amplitude $<$30\%
would have remained undetected in the \X\ light curves of \src.  

A final possibility, 
producing a relatively smooth modulation in the X--ray light curve,
is a self--eclipse of the hot emitting region
on the polar cap of a highly magnetic WD (polar)
in synchronous rotation with a low mass companion from
which it accretes matter via Roche-Lobe overflow 
without accretion disk formation.
In this case the profile and  duration of the eclipse
would be due to the shape and extension of the hot polar cap
periodically hidden at the WD spin period,
equal to the orbital period. The energy independence of the
eclipse depth would be also explained.

If \src\ were a Galactic CV as suggested by its low
X--ray luminosity for a 1\,kpc distance scale,
the folded light curve modulation would be clearly
more suggestive of an accretion stream eclipse or a self eclipse of the WD
in binary system of the polar class.


However,
evidence 
against the magnetic CV interpretation derives from the
analogy 
with the spectral characteristics observed from Polars and IPs. 
In both cases the observed spectra are compatible with a thermal 
bremsstrahlung spectrum with temperature in excess of tens of keV, 
plus (in many cases) a very soft spectral component
consistent with a blackbody of temperature in the 10-50\,eV range. 
The bolometric luminosity 
in this component is from one to several hundreds times higher than
the thermal bremsstrahlung component in Polars
(Ramsay \& Cropper \cite{ramcro03}), while it ranges from zero 
(i.e. the component is not detected) to several tens of times 
the thermal bremsstrahlung component in IPs. 

The spectral characteristics of \src\ are at variance with those 
of magnetic CVs.
The X--ray spectrum of \src\ has a characteristic 
temperature 2--3\,keV (that of the Comptonising electrons) and, 
in any case, cannot be fit with a thermal bremsstrahlung spectrum. 
The addition of a soft blackbody component with temperature
of 50 eV or less produces a bolometric blackbody flux comparable to
the bremsstrahlung, but the fit should be discarded
because of reduced $\chi^2$ of 1.6 (in the best case).

Finally we note that, in the Galactic interpretation, 
the X--ray luminosity  and spectrum of \src\ would be similar to 
those observed from  quiescent soft X--ray transient low mass binaries 
hosting a neutron star (see e.g. Campana \& Stella \cite{camste00}). 
X--ray eclipses have so far been observed only 
in a very low luminosity state of MXB\,1659-29 
(Wijnands \etal \cite{wijnands03}); 
no evidence for a dip--like modulation of the X-ray flux 
at the orbital period has yet been found in any quiescent 
neutron star soft X--ray transient. Therefore, while in principle viable, 
the possibility that \src\ is a dipping quiescent soft X--ray transient 
is not especially appealing. 
On the other hand the interpretation of he X-ray modulation of \src\
in terms of eclipses by the companion 
star is untenable because of the eclipse duration exceeds the maximum 
allowed duration (see above).


\subsection{Extragalactic scenario}
\label{extragalacticscen}

As an M31 source, the nearly steady luminosity of \src\
($\sim 10^{37}$\,erg\,s$^{-1}$ (0.3--10\,keV) would be similar
to that of moderately bright LMXRBs 
(White, Stella \& Parmar \cite{whitestella88}).
The shape of the 107\,min periodic modulation, comprising a fairly 
flat on-phase lasting for about 3/4 of the cycle and a relatively 
long and smooth dip (1/4 of the cycle) reaching (nearly) zero flux over 
some 1/10 of the cycle, is clearly reminiscent of X--ray dipping 
LMXRBs seen from high inclinations. 

The possibility that the observed modulation is due to true eclipses 
of a compact central X--ray source by the companion star (supported
by the square--wave fit of the folded \src\ light curves 
shown at the end of section \ref{res-timingan}),
can be simply ruled out because the observed eclipse duration 
is incompatible with
the geometry of the binary system (see section \ref{galacticscen}).

LMXRB dips are believed to arise from absorption and/or
Thomson scattering of X--rays coming from a central source by a bulge in the 
outer regions of the accretion disk, where the stream from the 
Roche-Lobe filling companion impacts the outer disk rim.  
The dipping phenomenon reflects the orbital period of the system, 
as demonstrated by a few systems which display both X--ray dips and eclipses
from the companion stars 
(e.g. MXB\,1659--298, Cominsky \& Wood \cite{comwood84}, 
Cominsky \& Wood \cite{comwood89}, 
Wijnands \etal \cite{wijnands03};
EXO\,0748--676, Parmar \etal \cite{parmar86};
GRS\,1747--312, into Zand \etal \cite{zand00}, in't Zand \etal \cite{zand03}).
Despite some jitter in phase and the common presence of pronounced
dip variability, the position of the bulge must remain 
(nearly) fixed in phase (Frank, King \& Lasota \cite{frakila87}). 
It is estimated that the dip phenomenon occurs in systems with an inclination
greater than $\sim70^{\circ}$.
Some 20 dipping LMXRBs are currently known in the Galaxy. 
Only one of these (X\,1916--053)
has an orbital period shorter than $\sim$2\,hr 
(White \& Swank \cite{whiteswank82}),
while two (4U\,1755--338 and 4U\,1630-47, in Mason \etal \cite{mason85} and 
Tomsik, Lapshow \& Kaaret \cite{tomsik98} respectively)  
likely host an accreting Black Hole Candidate (BHC).
Most dipping LMXRBs are type I X--ray bursters and thus contain an 
accreting neutron star.

In the dipping LMXRB interpretation \src\ would be a compact binary, 
the second dipper with a period below the period gap.

The spectra of high luminosity ($>$ $10^{37}$\,erg\,s$^{-1}$) 
LMXRBs hosting a central neutron star are 
usually well fit e.g. by a Comptonization model  
plus a blackbody component (White, Stella \& Parmar \cite{whitestella88}; 
for a review of other models see e.g. Di Salvo \& Stella \cite{disalvo02}).
When required, the blackbody component has a temperature 
around 1.5\,keV and it accounts for
$\sim$10--30\% of the total 1--10\,keV flux.  
For both low/high luminosity sources,
Comptonization parameters are $y \sim 2 - 4$, 
optical depths $\tau \sim 10-15$ 
and electron temperatures $T_e \sim 2 - 4$\,keV.
Spectra of (soft state) Black Hole Candidates in LMXRBs are well fit by 
Comptonization model too, but with a somewhat higher scattering depth 
($\tau > 15$) and lower electron temperature 
($T_e \sim$1\,keV; White, Stella \& Parmar \cite{whitestella88}).
LMXRBs with somewhat lower luminosity ($< 10^{37}$\,erg\,s$^{-1}$) 
are often dominated by a power law with photon index $\Gamma \simgt 2$ 
(Christian \& Swank \cite{chriswank97}; Schulz \cite{schulz99}; 
Church \& Ba{\l}uci\'{n}ska--Church \cite{churchbalu01}).

In the case of \src, the \X\ spectra are reasonably well fit 
by a power law or a Comptonization model (see Table \ref{specpar}).
The power-law fit gives a photon index ($\Gamma <1$), i.e. harder 
than that found in  LMXRBs.
The addition of a blackbody component to the power law, 
though not required from the statistical point of view,  
implies too high a blackbody temperature ($\sim 2.3$\,keV) 
and fraction of the total luminosity ($\sim 70$\%)
as compared to typical values for LMXRBs. 
On the contrary the thermal Comptonization model discussed in section 
\ref{res-spectralan} 
yields an electron temperature of $\sim 3$\,keV and an optical depth 
$\sim 13$ consistent with high luminosity LMXRBs hosting a neutron star.


In this interpretation other relevant  features of \src\ are

{\it (i)} the  $\sim$25\% duty cycle of the dips; 
this is in agreement with those observed in Galactic dipping sources:
the duty cycle of the dips varies from source to source and is 
typically 10\%--40\%.
Sources showing dips with a $\sim$20\% duty cycle are, for example, 
4U\,1755--338 (Mason \etal \cite{mason85}), 
XB\,1254--690 (Courvoisier \etal \cite{cour86}), 
MXB\,1659--29 (Cominsky \& Wood \cite{comwood84})

{\it (ii)} the shape of the dips, consistent with 
a relatively smooth average profile;
this is common in Galactic dipping sources as well, despite 
the fact that individual dips often display pronounced irregular 
variability. Owing to poor statistics, individual dip variability 
would not be revealed in \src.

{\it (iii)} the very high degree of modulation, nearly 100\%;  
in turn this is not unusual in LMXRBs: e.g. 
4U\,1916--053 and XB\,1254--690 often 
display more $>90$\% modulated dips; MXB\,1658--298 presented 
$\sim$90\% modulated dips in the \B\ observation of August 2000 
(Oosterbroek \etal \cite{ooster00}); the dips of
EXO\,0748--676 occasionally reach a depth of 80\% 
(Parmar \etal \cite{parmar86}). 

{\it (iv)} the absence of a conspicuous increase in $N_{\rm H}$ during dips.
The most constraining upper limit to the average absorption column 
of the obscuring bulge producing the dips 
we got from \X\ observations of \src\ is rather low: 
$< 2\times 10^{21}$\,cm$^{-2}$ (at 95\% confidence level; 
see Table \ref{specpar2}). 
However, we reiterate that the dip spectra were extracted 
over a fairly large phase interval (1/4 of a cycle), 
which samples predominantly the relatively shallow stages of the dip 
(this was done in order to increase the S/N). 
For this reason the upper limit we derived might underestimate 
the actual limit on the increase of $N_{\rm H}$ around dip minimum.

{\it (v)} the energy independence of the modulation 
(to within the uncertainties).
We have already discussed this point in section \ref{galacticscen} 
(5th paragraph).
As in the IP case, this may be accounted for if the disk bulge producing
the dips were ionized by the central X--ray source 
and the flux reduction mechanism at work 
was Thomson scattering and not (as often happens) photo-electric absorption.

Following the same line of reasoning of section  \ref{galacticscen} 
we conclude that for an X--ray luminosity 
$L_X \sim 1 \times 10^{37}$\,erg\,s$^{-1}$
the distance from the central X--ray source of the material causing the dips 
should be $R \simlt 3 \times 10^9$\,cm,
since only material closer than $3 \times 10^9$\,cm from 
the X--ray source will be significantly ionized. 
By contrast, for a 107\,min binary period 
the binary separation of the system is expected to be
$a \sim 5.2 \times 10^{10} M_X^{1/3} (1+q)^{1/3}$\,cm, 
with $M_X$ the mass of the compact object in solar units, 
and $q$ the mass ratio of the donor star and the accreting star.
Thus, the accretion disk is expected to have a radius of 
$\sim 0.3 a \simlt 2 \times 10^{10} M_X^{1/3}$\,cm  
for stellar masses with ratio $q \simlt 1$ 
(Whitehurst \& King \cite{whitehurst91}; Eggleton \cite{eggle83}).
This requires, for ionization to be important, that the disk is at least 
one order of magnitude smaller than expected or the material responsible
for the dips is located further inside the edge of the disk, 
both of which seem unlikely.
We note that also in the case of 4U\,1755--338 the ionization hypothesis 
leads to the same difficulties (Mason \etal \cite{mason85}).

Another possibility to interpret the energy independent dips of \src\ 
would be the eclipse by the companion star of an extended source of X--rays
of comparable size, such as that originating from electron scattering in 
an Accretion Disk Corona (ADC). Besides the energy-independence, 
this would explain 
the smooth ingress and egress of the 
dips as well as the (nearly) 100\% reduction of the observed X--ray flux.  
In this hypothesis, however the central X--ray source should be 
completely hidden by the accretion disk rim in order to explain the absence of 
sharp eclipses. 
In analogy with known ADC sources we would thus expect that the true 
luminosity of \src\ is one or two orders of magnitude higher than that 
observed, i.e. $10^{38}$--$10^{39}$\,erg\,s$^{-1}$. This in turn would suggest
a BHC accreting at very high rates; however the observed source spectrum is 
considerably harder than that expected from such a BHC. 
%


\section{Conclusion}
\label{conclusion}

We discovered 107~min periodic dips in the X--ray light curve of \src.
The dip profile and amplitude, and average spectra of \src\ 
are consistent with a accreting neutron star in a short period LMXRB 
seen at a high inclination and located in M31.
The apparent energy-independence of the dips is more difficult to interpret 
in this scenario.
If this interpretation was correct, \src\ would be
the second dipping source discovered in M31
(after XMMU\,J004314.1+410724 
in the globular cluster Bo\,158, Trudolyubov \etal \cite{trudo02}).

The possibility that \src\ is a foreground magnetic CV cannot be ruled 
out at present, but appears to be less likely.


\begin{acknowledgements}

We acknowledge a number of useful exchanges with F. Fiore, J. Osborne
and A. Tiengo on the analysis of \X\ data. 
The comments from an anonymous referee helped improving the analysis
and the presentation of the data.

This work was partially supported through ASI and MIUR--COFIN grants.

\end{acknowledgements}


\end{document}